\documentclass[prb,showpacs]{revtex4}

\usepackage{graphicx}
\usepackage{dcolumn}
\usepackage{bm}
\usepackage{epsfig}
\usepackage{pictex}

\begin{document}

\title{Dynamical $1/N$ approach to time-dependent currents through quantum dots}

\author{J. Merino}

\email{jaime.merino@uam.es}
\affiliation{Max-Planck-Institut f\"ur Festk\"orperforschung, D-70506 Stuttgart, Germany}
\altaffiliation[Current address:  ]{Dpt. de F\'isica Te\'orica de la Materia Condensada,
Universidad Aut\'onoma de Madrid, Spain}

\author{J. B. Marston}

\email{marston@physics.brown.edu}
\affiliation{Department of Physics, Brown University, Providence, RI 02912-1843 USA}

\date{\today}

\begin{abstract}
A systematic truncation of the many-body Hilbert space is
implemented to study how electrons in a quantum dot
attached to conducting leads respond to time-dependent biases.  The method,
which we call the dynamical $1/N$ approach,
is first tested in the most unfavorable case,
the case of spinless fermions ($N=1$).
We recover the expected behavior, including
transient ringing of the current in response to an abrupt change of bias.
We then apply the approach to the physical case of spinning electrons,
$N=2$, in the Kondo regime for the case of infinite intradot Coulomb
repulsion.  In agreement with previous calculations based on the non-crossing
approximation (NCA), we find current oscillations associated with transitions
between Kondo resonances situated at the Fermi levels of each lead.
We show that this behavior persists for a more realistic model of
semiconducting quantum dots in which the Coulomb repulsion is finite.
\end{abstract}


\maketitle
\bigskip

\section{INTRODUCTION}
\label{sec:intro}

The behavior of strongly interacting electrons confined to low spatial dimensions
and driven out of equilibrium is still poorly understood.
Recent advances in the construction of small quantum dot devices have opened up
the possibility of studying in a controlled way the nonequilibrium
behavior of strongly correlated electrons.  For
instance the Kondo effect, which was first observed in metals
with dilute magnetic impurities\cite{Hewson}, has now been seen in
measurements of the conductance through a single-electron
transistor\cite{Goldhaber} (SET), in accord with theoretical
predictions\cite{Lee}.  Most experiments so far
have focused on steady state transport through a quantum dot.

Several theoretical approaches have been developed to calculate
electrical transport properties in a far from equilibrium situation.
Currents generated by a large static bias applied to the leads of a
quantum dot in the Kondo regime have been analyzed in some detail
using the Keldysh formalism combined with the
non-crossing approximation (NCA)\cite{Wingreen}.  Remarkably, stationary
currents may also be found exactly with the use of the
Bethe ansatz\cite{Ludwig}.  Exact treatments at special points
in parameter space have also been carried out\cite{Schiller}.
More recently, attention has been paid to time-dependent phenomena in
quantum many-body systems\cite{Wingreenb,Schiller,Ng,Nordlander,
Plihal,Kaminski99,Talyanskii97,Cuniberti98}.
By looking at the response of an interacting dot to
a time-dependent potential, Plihal, Langreth, and Nordlander
studied the several time scales associated with
different electronic processes\cite{Nordlander}.
The response to a sinusoidal AC potential\cite{Kouwenhoven,Lopez,Kaminski99} in
the Kondo regime has been studied both in the
low\cite{Ng} and high-frequency limits\cite{Hettler}.

NCA has been the most commonly used approximation to obtain
response currents due to an external applied pulse.
The exact solution at the Toulouse point corroborates many of
the NCA results\cite{Schiller}.  However, NCA calculations have mostly
focused on the Kondo regime at not too low temperatures, because in the mixed-valence
regime, or at low temperatures, NCA is known to give spurious results
\cite{Costib}.  In practice NCA has usually been limited to studies in
the $U \rightarrow \infty$ limit.  For a typical SET, the ratio of the Coulomb
repulsion in the dot to the dot-lead coupling strength ranges over
$U/\Delta \approx 10$ to $20$.  Other numerical techniques, such as the Numerical
Renormalization Group, have also been
used\cite{Costi} to calculate the conductance through dots. However,
these methods seem to be limited to the static linear-response regime.
Perturbative RG methods \cite{KonigRG} have been used
to analyze nonequilibrium transport through dots in the Kondo
regime\cite{Roschb}.  Recent progress in extending density matrix
renormalization group (DMRG) methods to explicitly time-dependent
problems has been recently achieved with the Time-dependent
Density-Matrix Renormalization-Group (TDMRG)\cite{Cazalilla}.

In this paper we introduce a dynamical $1/N$ approach, where
$N$ counts the number of spin components of the electron.  Physical
electrons correspond to $N = 2$ (spin-up or spin-down).  We will also
have occasion to consider spinless electrons ($N = 1$) as this
case permits a comparison with known exact results for
noninteracting electrons.  Higher values of $N > 2$ are of actual
physical interest too, as these can occur when there are additional orbital
and channel degeneracies.  The model we study possesses a
global SU(N) spin symmetry.

Unlike NCA, the dynamical
$1/N$ approach is systematic in the sense that it includes all Feynman
diagrams up through a given order in $1/N$; not only a certain class of them.
As crossing diagrams are neglected in the NCA it is of
interest to compare the two approaches.  
The static version of the $1/N$ expansion
is a type of configuration-interaction (CI) expansion of the sort familiar
to quantum chemists.
The dynamical 1/N approach has some similarities to a real-time perturbation
scheme along the Keldysh contour developed by K\"onig\cite{Konig}.
Perturbing in the powers of the lead-dot coupling
generates an increasing number of particle-hole excitations in the leads.
In practice a resonant-tunneling approximation is implemented to restrict the
set of diagrams to be considered.  Only single particle-hole excitations
in the leads were included in the off-diagonal elements of the total density matrix,
and the intra-dot Coulomb repulsion $U$ was taken to be infinite.

We determine the response currents through a quantum dot under the influence of
both a small step bias, and a large pulse. Transient oscillatory phenomena
are found. In particular, a type of ringing in the current discovered previously
for non-interacting electrons is recovered within
the $1/N$ approach upon setting $N=1$.  We compare the behavior
of the response current for spinning and spinless electrons,
and demonstrate that the intradot Coulomb interaction changes both the period
of the oscillations and the decay rate of the currents.  We further demonstrate that
the period of the oscillations, in the case of spinning electrons, does not depend
on the dot level energy when the dot level is moved
from the Kondo into the mixed valence regime, and thus extend previous
results in the Kondo regime based on NCA\cite{Plihal}.

The paper is organized as follows: In Section \ref{sec:model} we
introduce a generalized time-dependent Newns-Anderson model for
describing a quantum dot and its systematic solution using a $1/N$
expansion of the many-body wavefunction.  Also we describe how non-zero
temperature can be treated in the method and how observables
are calculated. In Section \ref{sec:step} we find the
current through a quantum dot for a small symmetric bias,
deep in the linear response regime. In Section \ref{sec:pulse}
we analyze the response of a quantum dot to a
large amplitude bias for spinless and interacting electrons.
We conclude in Section \ref{sec:concl} by discussing the advantages
and limitations of the dynamical $1/N$ approach, as well as its future prospects.
Details of the equations of motion and the calculation of currents
are presented in an Appendix.

\section{Theoretical approach}
\label{sec:model}

In this section we discuss the correlated-electron model that we
employ to study time-dependent currents in a quantum dot.  We then
present its systematic solution order by order in powers of $1/N$.

\subsection{Generalized Newns-Anderson Model}
\begin{figure}
\includegraphics[width=18cm,height=8cm]{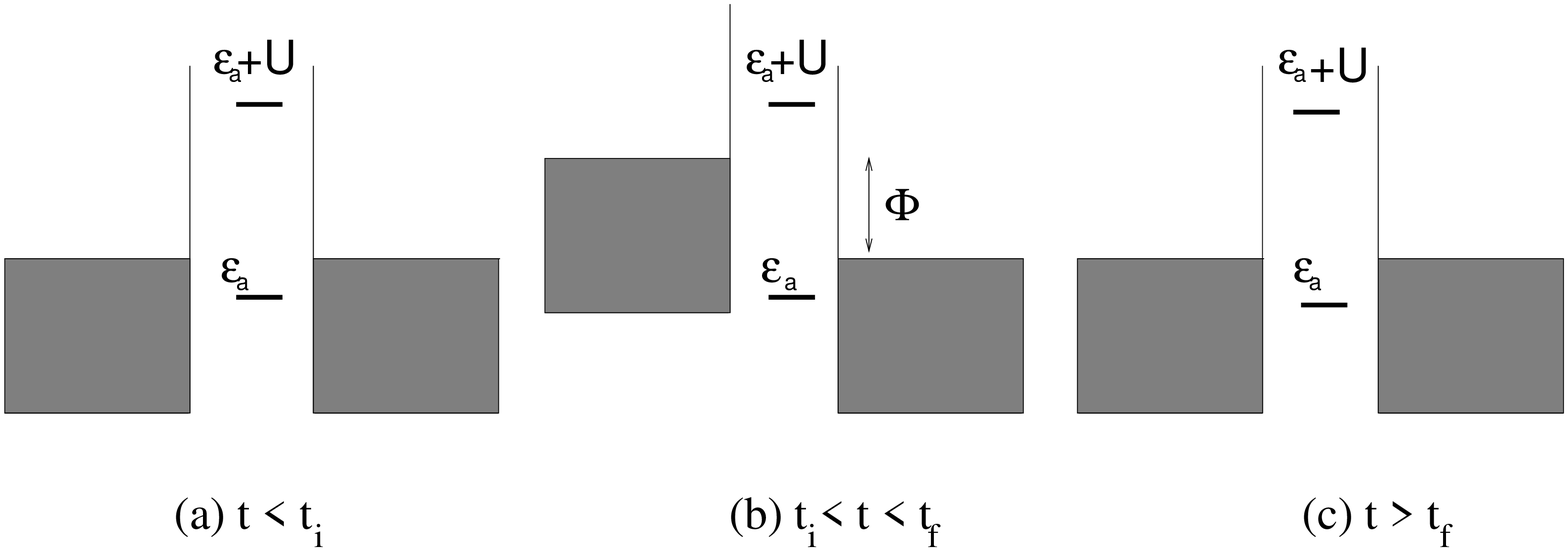}
\caption{\label{fig1}Energy level representation of interacting electrons
in a quantum dot before, during, and after application of a bias pulse
of amplitude $\Phi$ to the left lead,
holding both the right lead chemical potential and the dot energy level,
$\epsilon_a$, fixed.  Here $U$ is the Coulomb
interaction between electrons inside the dot.
}
\end{figure}

The model we use to analyze transport through a
quantum dot is defined by the following generalized time-dependent
Newns-Anderson Hamiltonian:

\begin{eqnarray}
H(t) &=& \sum_a [\epsilon^{(1)}_a(t) \hat P_1  + \epsilon^{(2)}_a(t) \hat P_2]~
\ c_a^{\dag \sigma} c_{a \sigma}
 +  \sum_{k \alpha} \epsilon_{k \alpha}(t) \ c_{k \alpha}^{\dag \sigma} c_{k \alpha \sigma}
\nonumber \\
&+& {{1}\over{\sqrt{N}}}~ \sum_{a;\ k \alpha} \big{\{} [V^{(1)}_{a;k \alpha}(t) \hat P_1
+ V^{(2)}_{a;k \alpha}(t) \hat P_2]
\ c_a^{\dag \sigma} c_{k \alpha \sigma} + H.c \big{\}}
\nonumber \\
&+&  {{1}\over{2}} ~ \sum_a U_{aa} n_a (n_a - 1)
+ \sum_{a > b} U_{ab} n_a n_b ~.
\label{AN}
\end{eqnarray}
Here $c_a^{\dag \sigma}$ is an operator that creates an electron in the
dot level $a$ with spin $\sigma$ and $c_{k \alpha}^{\dag \sigma}$
creates an electron in a level $k$ in the $\alpha$-lead.
Throughout the paper we adopt the following notation for the indices:  Greek
letters $\alpha$, $\beta$, $\gamma$ and $\delta$ label the left and right leads.
Greek letter $\sigma$ labels the spin, and a sum over the $N$ spin components
is implied whenever there is a pair of raised and lowered indices.
Operator $n_a \equiv c_a^{\dag \sigma} c_{a \sigma}$ counts the occupancy of level
$a$ in the dot.  Projection operators $\hat P_1$ and $\hat P_2$ can be written
in terms of $n_a$ and project onto the
single- and double-occupied dot, respectively. Parameters
$\epsilon_a^{(1)}(t)$, $\epsilon_a^{(2)}(t)$, $V_{a;k}^{(1)}(t)$
and $V_{a;k}^{(2)}(t)$ are respectively the orbital energies and hybridization
matrix elements for the quantum dot with one or two electrons.
The projection operators enable the use
of different couplings and orbital energies depending on the number
of electrons in the dot. For simplicity, however, here we assume
that the hybridization matrix elements are identical, and
independent of level index and time:
$V_{a;k \alpha}^{(1)}(t)=V_{a;k \alpha}^{(2)}(t) \equiv V$.
We also assume that the electronic levels in the dot
are the same (apart from the charging energy $U_{aa}$): $\epsilon^{(1)}_a(t)
= \epsilon^{(2)}_a(t) = \epsilon_a(t)$.  In Eq. \ref{AN},
the original hybridization matrix elements are divided by
a factor of $\sqrt{N}$ so that the dot level
half-width due to the hybridization with either lead,
$\Delta_{\alpha} = \pi \rho |V|^2 = \Delta$,
is independent of $N$, remaining finite in the
$N \rightarrow \infty$ limit. The density of states per spin
channel of either lead is denoted by $\rho$, here is taken to be
constant. Coulomb interaction $U_{aa}$ is the
repulsive energy between two electrons that occupy the same dot level
$a$, while $U_{ab}$ is the interaction between electrons in different levels.
Time-dependence may come in through the conducting leads,
$\epsilon_{k \alpha}(t) = \epsilon_{k \alpha} + \Phi_{k
\alpha}(t)$, the dot level, $\epsilon_a(t)= \epsilon_a + \Phi_a(t)$,
or both.  Functions $\Phi_{a}(t)$ and $\Phi_{k \alpha}(t)$ may have
any time dependence.  For instance Fig. \ref{fig1} illustrates the
case of a rectangular pulse bias potential applied to the left
lead keeping both the dot level and right lead unchanged.

To permit a comparison of the dynamical $1/N$ approach to other methods
we further simplify the above Hamiltonian.  In the following
we only consider the case of a single level, $a = 0$, so for notational simplicity
we define $U_{aa} \equiv U$ in the rest of the paper.  The model then corresponds to
the usual one considered by many other authors that treats only
a single s-wave level in the quantum dot.  However, we stress that the
Hamiltonian, Eq. \ref{AN}, can also be used to study more general
and experimentally relevant models.  

\subsection{$1/N$ expansion of the wavefunction}
The many-body wavefunction is constructed by systematically expanding
the Hilbert space into sectors with increasing number of
particle-hole pairs in the leads. Sectors with more and more particle-hole pairs are
reduced by powers of $1/N$ in the expansion. The approach was originally
introduced to study magnetic impurities
in metals\cite{Varma}, mixed-valence compounds\cite{Gunnarsson}
and it was first applied to
a dynamical atom-surface scattering problem by Brako and Newns\cite{Brako}.
We have previously applied it to the scattering of alkali\cite{Ernie,Onufriev} and
alkaline-earth ions such as calcium off metal surfaces\cite{Merino} where in the
latter case various Kondo effects may be expected\cite{SNL,Toulouse}.
The expansion of the time-dependent wavefunction for the lead-dot-lead system
up to order O($1/N^2$) in the spin-singlet (more generally, SU(N)-singlet)
sector may be written in terms of the scalar amplitudes $f(t)$,
$b_{a;k \alpha}(t)$, $e_{L \gamma,k \alpha}(t)$, $d_{k \alpha,q \beta}(t)$,
$s_{a; L\gamma,k\alpha,q\beta}(t)$, $a_{a; L\gamma,k\alpha,q\beta}(t)$,
$g_{L\gamma,P\delta,k\alpha ,q\beta}(t)$, and $h_{L\gamma,P\delta,k\alpha,q\beta}(t)$:
\begin{eqnarray}
| \Psi(t) \rangle &=& f(t)~ | 0 \rangle
+ \sum_{a;\ k, \alpha} b_{a;k \alpha}(t)~ |a; k \alpha \rangle
+ \sum_{L,\ k, \gamma, \alpha} e_{L \gamma,k \alpha}(t)~ |L \gamma, k \alpha\rangle
+ \sum_{q<k, \alpha, \beta} d_{k \alpha,q \beta}(t)~ |k \alpha, q \beta \rangle
\nonumber \\
&+& \sum_{a;\ L,\ q <k, \gamma, \alpha, \beta } s_{a; L\gamma,k\alpha,q\beta}(t)~
{|a; L\gamma, k\alpha, q\beta \rangle}^{S}
+ \sum_{a;\ L,\ q<k, \gamma, \alpha, \beta} a_{a; L\gamma,k\alpha,q\beta}(t)~
{|a; L\gamma,k\alpha, q\beta \rangle}^{A}
\nonumber \\
&+& \sum_{L>P,\ q<k, \gamma, \delta, \alpha, \beta} g_{L\gamma,P\delta,k\alpha ,q\beta}(t)~
{|L\gamma, P\delta, k\alpha, q\beta \rangle}^{S}
+ \sum_{L>P,\ q<k,\gamma,\delta, \alpha, \beta} h_{L\gamma,P\delta,k\alpha,q\beta}(t)~
{|L\gamma, P\delta, k\alpha, q\beta \rangle}^{A}
\nonumber \\
&+& \{{\rm rest~ of~ Hilbert~ space} \}\ .
\label{O.PSI}
\end{eqnarray}
The first line of this equation contains the O(1) amplitudes, the second line has the
O($1/N$) terms, and the amplitudes in the third line are O($1/N^2$).
Upper case Roman letters $L$ and $P$ label lead levels above the Fermi level,
whereas lower case letters $k$ and $q$ label lead levels below
the Fermi energy. Expressions for the basis states in
the case of atoms scattered off metal surfaces can be found elsewhere\cite{Onufriev},
but we rewrite them here for clarity.  The vacuum state $| 0 \rangle$ represents an
empty dot, with both leads filled with electrons up to the Fermi energy.  The remaining states,
which are all SU(N)-singlets due to the contracted spin indices $\sigma$ and $\sigma^\prime$, are:
\begin{eqnarray}
|a; k \alpha \rangle &=&
{{1}\over{\sqrt{N}}}~ c^{\dagger \sigma}_a c_{k \alpha \sigma} |0 \rangle
\nonumber \\
|L \gamma, k \alpha\rangle &=&
{{1}\over{\sqrt{N}}}~ c^{\dagger \sigma}_{L \gamma} c_{k \alpha \sigma} |0 \rangle
\nonumber \\
|k \alpha, q \beta \rangle &=&
{{1}\over{\sqrt{N (N-1)}}}~ c^{\dagger \sigma}_0 c_{k \alpha \sigma}
c_0^{\dagger \sigma^\prime} c_{q \beta \sigma^\prime} |0 \rangle
\nonumber \\
{|a; L\gamma, k\alpha, q\beta \rangle}^{S} &=&
{{1}\over{\sqrt{2N (N-1)}}}~ \big{\{} c_{L \gamma}^{\dagger \sigma} c_{k \alpha \sigma}
c^{\dagger \sigma^\prime }_a c_{q \beta \sigma^\prime} |0 \rangle
+ c_{L \gamma}^{\dagger \sigma} c_{q \beta \sigma }
c^{\dagger \sigma^\prime}_a c_{k \alpha \sigma^\prime} |0 \rangle \big{\}}
\nonumber \\
{|a; L\gamma, k\alpha, q\beta \rangle}^{A} &=&
{{1}\over{\sqrt{2N (N+1)}}}~ \big{\{} c_{L \gamma}^{\dagger \sigma} c_{k \alpha \sigma}
c^{\dagger \sigma^\prime}_a c_{q \beta \sigma^\prime} |0 \rangle
- c_{L \gamma}^{\dagger \sigma} c_{q \beta \sigma }
c^{\dagger \sigma^\prime}_a c_{k \alpha \sigma^\prime} |0 \rangle \big{\}}
\nonumber \\
{|L\gamma, P\delta, k\alpha, q\beta \rangle}^{S} &=&
{{1}\over{\sqrt{2N (N-1)}}}~ \big{\{} c_{L \gamma}^{\dagger \sigma} c_{k \alpha \sigma}
c^{\dagger \sigma^\prime}_{P \delta} c_{q \beta \sigma^\prime} |0 \rangle
+ c_{L \gamma}^{\dagger \sigma} c_{q \beta \sigma }
c^{\dagger \sigma^\prime}_{P \delta} c_{k \alpha \sigma^\prime} |0 \rangle \big{\}}
\nonumber \\
{|L\gamma, P\delta, k\alpha, q\beta \rangle}^{A} &=&
{{1}\over{\sqrt{2N (N+1)}}}~ \big{\{} c_{L \gamma}^{\dagger \sigma} c_{k \alpha \sigma}
c^{\dagger \sigma^\prime}_{P \delta} c_{q \beta \sigma^\prime} |0 \rangle
- c_{L \gamma}^{\dagger \sigma} c_{q \beta \sigma }
c^{\dagger \sigma^\prime}_{P \delta} c_{k \alpha \sigma^\prime} |0 \rangle \big{\}}.
\nonumber \\
\end{eqnarray}
We set the Fermi energy to zero ($\epsilon_F \equiv 0$) before the pulse is applied ($t<t_i$).
An ordering convention is imposed on the above states such that $\epsilon_q < \epsilon_k < 0$ and
$\epsilon_L > \epsilon_P > 0$.  Fig. \ref{fig2} shows a schematic representation
of the different Hilbert space sectors appearing in the $1/N$ expansion up through $O(1/N^2)$.
Each row represents a different order in the $1/N$ expansion,
increasing upon moving downwards in the diagram.
The physical interpretation of each of the sectors is as follows:
State $| L \gamma, k \alpha \rangle$ represents an electron excited to level
$L$ in lead $\gamma$ along with a hole in level $k$ in lead $\alpha$.
State $| k \alpha, q \beta \rangle $ represents configurations in which two
electrons occupy the lowest level ($a = 0$) of the quantum dot simultaneously,
with two holes left behind in the leads.
These configurations are suppressed in the limit $U \rightarrow \infty $. 
States ${|a; L \gamma, k \alpha, q \beta \rangle}^{S}$ and ${|a; L\gamma,
k \alpha, q \beta \rangle}^{A}$ are symmetric (S) and
antisymmetric (A) combinations of the configuration with one
electron in level $L$ and holes in level $k$ of lead $\alpha$
and in level $q$ of lead $\beta$.  The division of the sector into two parts
reflects the fact that the state produced by an electron hopping to the dot
from a continuum level $k$ while another electron is excited from
$q$ to $L$ can be distinguished (because the electrons carry spin)
from the state in which $k$ and $q$ are interchanged.
A similar decomposition is carried out
for the sectors with two particle-hole pairs described by amplitudes $g$ and $h$.  

We neglect, at O($1/N^2$), the sector
corresponding to a singly-occupied dot with a hole and two particle-hole pairs
as the amplitude for this sector has 5 continuum indices.  This sector 
is expected to be important for the physically interesting case of the 
Kondo and mixed-valent regimes with near single occupancy of the quantum dot. 
Its inclusion would presumably improve the behavior of the dynamical $1/N$ approach at long
times, and we leave this for future work. 
Strictly speaking, through O($1/N^2$) we should also include sectors corresponding to
a doubly-occupied quantum dot with one or two particle-hole pairs in addition
to the two holes in the leads.  However,
as these configurations are described by amplitudes with up to 
6 different continuum
indices, and as the amplitudes for these sectors is small (due to the repulsive
Coulomb interaction), the computational work required to include the
amplitudes is excessive, and we drop them from the equations of motion.

The equations of motion for the amplitudes that appear in Eq. \ref{O.PSI} are given in the
Appendix.  We note that the terms we keep in the $1/N$ expansion is equivalent to summing
up Feynman diagrams, including the crossing ones, up to order $1/N^2$.  The inclusion of
crossing diagrams is significant because it is these diagrams that are known to
be responsible for the recovery of Fermi liquid behavior at low temperatures\cite{Bickers}, 
and the disappearance of the Kondo effect in the spinless $N \rightarrow 1$ limit.

\subsection{Calculation of observable quantities}
We conclude our discussion of the dynamical $1/N$ method by explaining
how observable quantities such as the currents are
calculated. At zero temperatures, the initial state of the system
prior to application of the bias is chosen to be the ground state,
which is obtained by the power method.  Integrating the equations of motion
forward in time then yields $| \Psi(t) \rangle$, and expectation values of
observables, $\hat{O}$, are calculated periodically during the course of the time
evolution:
\begin{equation}
\langle \hat{O}(t) \rangle= \langle \Psi(t) | \hat{O} | \Psi(t) \rangle\ .
\end{equation}
For the particulars of how to calculate the expectation value of the current
operators, see the Appendix.

Most of the results presented in the paper, with the exception of those shown
in the final figure, are for the case of zero temperature.
At non-zero temperatures we must extend the method\cite{Merino}.
Suppose that at an initial time $t = 0$, prior to application of the bias,
we could find the complete set of energy eigenstates and values
$\{ |\Psi_n(0) \rangle,~ E_n \}$ satisfying $H | \Psi_n(0) \rangle
= E_n | \Psi_n(0) \rangle$.  Each of these eigenstates could then
be evolved forward in time yielding $\{ | \Psi_n(t) \rangle \} $.  The
combined thermal and quantum average of any observable quantity, $\hat O$,
would then be given at time $t$ by:
\begin{equation}
\langle \hat{O}(t) \rangle = {{1}\over{Z}} \sum_n \langle \Psi_n(t)|
\hat{O} | \Psi_n(t) \rangle \exp \bigg{\{} -{{E_n}\over{k_B T}} \bigg{\}}
\end{equation}
where
\begin{equation}
Z = \sum_n \exp \bigg{\{} -{{E_n}\over{k_B T}} \bigg{\}}
\end{equation}
is as usual the partition function of the system.  However, as
the Hilbert space is enormous in size, the determination
of the whole set of time-evolved many-body wavefunctions is
prohibitively difficult.  Instead we sample the
Hilbert space by creating a finite, random, set of wavefunctions:
$\{ | \Xi_n \rangle,~ n = 1, \ldots, n_{sample} \}$.
The wavefunctions are generated by assigning random numbers to
the amplitudes that describe the different sectors of the wavefunction,
Eq. \ref{O.PSI}.  Each of these randomly generated wavefunctions
is then weighted by half of the usual Boltzmann factor,
\begin{equation}
| \tilde{\Psi}_n(0) \rangle = \exp \bigg{\{} -{{H(0)}\over{2
k_B T}} \bigg{\}} | \Xi_n \rangle\ ,
\label{Boltz}
\end{equation}
and the resulting set of weighted wavefunctions is time-evolved to
yield $ \{ | \tilde{\Psi}_n(t) \rangle \} $.
(The wavefunctions are weighted by half the usual Boltzmann factor so that
thermal expectation values, 
which depend on the square of the wavefunctions, have the correct weight.)
Thermal averages of each observable such as the current are calculated
periodically during the course of the integration forward in time:
\begin{equation}
\langle \hat{O}(t) \rangle= {{1}\over{Z}} \sum_{n=1}^{n_{sample}} \langle
\tilde{\Psi}_n(t) | \hat{O} | \tilde{\Psi}_n (t) \rangle
\label{aver}
\end{equation}
where
\begin{equation}
Z = \sum_{n=1}^{n_{sample}} \langle \tilde{\Psi}_n(t) | \tilde{\Psi}_n(t) \rangle
= \sum_{n=1}^{n_{sample}} \langle \tilde{\Psi}_n(0) | \tilde{\Psi}_n(0) \rangle\ .
\end{equation}
At the very lowest temperatures only a single sample is needed as the approach
then reduces to the zero-temperature power method described above.
At non-zero but moderate temperatures, thermal averages over $n_{sample} = 30$
random wavefunctions suffice, and the convergence of the
thermal average may be easily checked by increasing the number of samples.
We stress that time dependence comes in only via the
wavefunctions.  This makes the present approach rather
straightforward, avoiding cumbersome Green's function formalisms
based on the Keldysh or Kadanoff-Baym techniques. On the other
hand, the approach is not useful at very high temperatures as an
excessive number of samples is then required.

We now present an intuitive picture of how electronic transitions
take place as the current flows through the quantum dot.
Fig. \ref{fig2} shows some of the sectors,
up to O(1/$N^2$), that contribute to the current through a dot
subject to the rectangular pulse bias shown in Fig. \ref{fig1}.
When the bias is switched on, electronic transitions between the dot
and the leads take place. Sectors shown in Fig. \ref{fig2} in which
an electron travels from the left to the right leads creating cross-lead
particle-hole excitations make the most important contribution to the current.
The current through the dot is a consequence of the formation of
particle-hole pairs, with holes accumulating in one lead and particles in the other.
Sectors of the Hilbert space with increasing numbers of particle-hole pairs,
and hence higher order in $1/N$, become populated as time goes on.
Electronic transitions back
down to lower order sectors also occur, of course, since the
Hamiltonian is Hermitian.  However, because the phase space of sectors
with increasing numbers of particle-hole pairs grows rapidly with
the number of pairs, when the system
is driven out-of-equilibrium, reverse processes back down to
lower orders occur less frequently.  This irreversibility may be
quantified in terms of an increasing entropy\cite{Onufriev}.

\section{Response to a small, symmetric, step in the bias potential}
\label{sec:step}

We first apply the dynamical $1/N$ approach to study the response
of a quantum dot to a small, symmetric, step bias potential.
We use parameters appropriate for a semiconducting quantum dot. We
either take $U \rightarrow \infty$ or, more realistically,
set $U = 1$ or $2$ meV comparable to the values reported by
Goldhaber-Gordon {\it et al.} for the single electron transistor
built on the surface of a GaAs/AlGaAs heterostructure that led
to the reported Kondo effect\cite{Goldhaber,Gores}.
We further set the dot-lead half-width to be $\Delta = 0.4$ meV, and
$\epsilon_a = -2 \Delta$.  Thus in the case of spinning electrons
the dot is in the Kondo regime.  In a real SET there
are several quantized levels with a typical spacing of order
$0.4$ meV.  However, as mentioned in the previous section,
here for simplicity we consider only the case of a single energy level in the dot.
At time $t_i = 0.5 \hbar/\Delta$ the bias on the left lead
is suddenly turned on, raising the Fermi energy by $\Phi/2$, while
the right lead is shifted down by the opposite amount, -$\Phi/2$,
and the dot level is held fixed. The size of the bias is chosen to be
$\Phi = 0.05 \Delta$ which is small
enough to induce a linear response in the current.

The leads are assumed to be described by a flat band of constant
density of states with all energies taken to be relative to the
Fermi energy. The band is taken to be symmetrical about the Fermi
energy with a half-bandwidth set at $D = 4$ meV. We use $M = 30$
discrete levels both above, and below, the Fermi energy in our
calculations, except for the O($1/N^2$) amplitudes
$g_{L\gamma,P\delta,k\alpha ,q\beta}(t)$ and
$h_{L\gamma,P\delta,k\alpha,q\beta}(t)$. As these amplitudes each
have four continuum indices, we retain only 10 levels above, and
below, the Fermi energy.  This turns out to suffice as there is
little change when only 5 levels are retained.  To improve the
discrete description, the continuum of electronic states in the
conducting leads is sampled unevenly: the mesh is made finer near
the Fermi energy to account for particle-hole excitations of low
energy.  Specifically, the discrete energy levels below the Fermi
energy are of the following form:
\begin{equation}
\epsilon_k = -{{D}\over{e^\gamma - 1}}~ [e^{\gamma (k - 1/2) /
M} - 1],\ \ \ k = 1, 2, \cdots, M\
\end{equation}
with a similar equation for levels above the Fermi energy.
Thus, the spacing of the energy levels close to the Fermi energy
is reduced from the evenly spaced energy interval of $D/M$ by a factor of
$\gamma / (e^\gamma - 1) \approx 1/13$ for a typical choice of the sampling
parameter $\gamma = 4$.  The matrix elements $V^{(i)}_{a; k \alpha}$ are
likewise adjusted to compensate for the uneven sampling of the conducting
levels\cite{Merino}.
We note that the discretization of the continuum of electronic states
in the leads means that we can only study pulses of duration less
than a cutoff time: $t_{cutoff} \approx  2 \pi \hbar / \delta$,
where $\delta \approx D/(13 M)$ is the
level spacing close to the Fermi energy of the conducting leads.
As we show below, however, the $1/N$ expansion itself imposes a more
severe restriction on the reliability of the method at long times.

\subsection{Noninteracting spinless electrons ($N = 1$)}
The case of spinless electrons deserves special attention as it
provides a stringent test of the $1/N$ expansion, and is also
exactly solvable. For an Anderson impurity model at
equilibrium it is known\cite{Gunnarsson} that the occupancy of the
impurity at $N = 1$ is accurate to within 1\% of the exact
result when terms in the wavefunction expansion
are kept up through order $1/N^2$.  In the dynamical case  
shown in Fig. \ref{fig3}, however, we find that for $\epsilon_a = -2 \Delta$ the  
currents decay in time despite the fact that the bias remains turned on.
Insight into the breakdown of the dynamical $1/N$ expansion can be gleaned from 
the equilibrium problem as discussed by Gunnarssson and Sch\"onhammer\cite{Gunnarsson}.
They observe that the parameter range $\epsilon_a \ll -\Delta$ is an unfavorable 
one for the $1/N$ expansion.  The O($1$) approximation in this limit 
differs qualitatively from the O($1/N$) and O($1/N^2$) approximations. 
In particular there is a large change in variational ground state energy 
(97\% in the case of $\epsilon_a = -\Delta$) going from the O($1$) to O($1/N$) approximations.  
This compares to a change of only 18\% in energy for the $\epsilon_a = 0$ case.   
In the dynamical problem the observation likewise suggests that the case 
$\epsilon_a = -2 \Delta$ depicted in Fig. \ref{fig3} is also unfavorable for the 
$1/N$ expansion as higher order sectors may be expected to contribute substantially 
to the wavefunction.  By contrast the other $N = 1$ case of $\epsilon_a = 0$ shown 
in Fig. \ref{fig3} exhibits a clear plateau in the currents.  We interpret this 
behavior as follows:  As the dot energy level 
is raised from $\epsilon_a = -2 \Delta$ to $0$, the many-body wavefunction moves closer 
to the beginning ``root'' state of the Hilbert space expansion, 
the state $| 0 \rangle$ in Fig. \ref{fig2} for which
the dot level is empty. Higher-order terms in the $1/N$ expansion are less
important in this limit, and the expansion is more accurate.

The limited number of particle-hole pairs retained in the $1/N$
expansion means that electron transfer from one lead to the other
ceases when the higher order sectors become significantly
populated.  Therefore, we conclude that the dynamical $1/N$ 
approach is limited to rather short times after the step in the
bias is applied.  Inclusion of higher-order sectors in the
$1/N$ expansion would presumably permit accurate time-evolution
for longer times.  The amplitude of the $O(1/N^2)$ sector
for the case of Fig. \ref{fig3} is quite small when
the current starts to decay, consistent with the observation that 
the amplitude of the empty dot sectors (first column in Fig. \ref{fig2}) 
are small in comparison to the singly-occupied sectors (second column of Fig.
\ref{fig2}), in the the Kondo and mixed-valence regimes. 
Including the O($1/N^2$) sector consisting of two
particle-holes in the leads, an electron in the dot and an extra
hole in the leads should improve the behavior of the
current at longer times.  However as a practical matter the rapid
growth in computational complexity at high orders in the $1/N$ 
expansion prohibits extensions to
arbitrarily high order.  Cluster expansions based upon
exponentials of particle-hole creation operators\cite{Sebastian}
may offer a way to obtain more satisfactory behavior at long times.

\subsection{Interacting spinning electrons ($N = 2$)}
Physical electrons have spin, and since $N = 2$ is a more favorable case
from the standpoint of the $1/N$ expansion, we expect (and find) improved
behavior. The Coulomb repulsion $U$ reduces the size of higher-order
corrections as it acts in part to suppress charge-transfer
through the dot that leads to the formation of the particle-hole pairs.  
Furthermore the Kondo effect in the limit $\epsilon_a \ll -\Delta$ is 
recovered already at O($1$) in the $1/N$ expansion.  
As shown in Fig. \ref{fig3}, for $U \rightarrow \infty$ there is
a strong enhancement in the current in comparison to the spinless case ($N=1$).
This is as expected from the increase of spectral weight at the Fermi
level of the leads due to the Kondo resonance.  The current is
further enhanced at finite $U = 1$ meV.  This can
be understood from $1/N$ calculations of the spectral density at
equilibrium.  A decrease in $U$ results in an
increase in the spectral weight at the Fermi energy, reflecting a
corresponding increase in the Kondo energy scale (see Fig. 9 of Ref.
\onlinecite{Gunnarsson}).  The point is also illustrated by
an analytical estimate of the Kondo temperature for finite $U$
based on weak-coupling renormalization-group scaling\cite{Haldane}:
\begin{equation}
k_B T_K \approx \sqrt{{{\Delta_{dot} U}\over{2}}}~
\exp{ \bigg{\{} {{\pi \epsilon_a (\epsilon_a + U)}\over{\Delta_{dot} U}} }
\bigg{\}}\ .
\label{Hal}
\end{equation}
Here $\Delta_{dot} = \Delta_L + \Delta_R = 2 \Delta$ is the
half-width of the dot level due to its hybridization with both leads.
This formula yields $T_K \approx 4$K.  As the weak-coupling expression
Eq. \ref{Hal} is only reliable at small values of $U$, it likely
overestimates the Kondo temperature in the present case of $U/\Delta
= 2.5$.  It is worth comparing the Kondo temperature so obtained with the value
extracted from the $U \rightarrow \infty$ Bethe-ansatz expression\cite{Hewson},
\begin{equation}
k_B T_K = D {{e^{1+C-{{3}\over{2 N}}}}\over{2 \pi}} \bigg{(}{\Delta_{dot}\over{\pi D}}
\bigg{)}^{1/N} \exp{\bigg{\{} {{\pi \epsilon_a}\over{\Delta_{dot}} } } \bigg{\}}\ ,
\label{BA}
\end{equation}
where $C=0.577216$ is Euler's constant. Taking $N=2$, Eq. \ref{BA}
gives only $T_K \approx 185$ mK.  

Hence, the increase in the Kondo temperature
obtained from Eqs. \ref{BA} and \ref{Hal} as $U$ is decreased from infinite to finite values is 
consistent with the enhancement of the current shown in Fig. \ref{fig3}. 
We use Eq. \ref{Hal} to estimate the Kondo temperature at finite-$U$ in 
the following section. 

\section{Response to a large rectangular pulse bias potential}
\label{sec:pulse}

In this section we study the response of the
quantum lead-dot-lead system to a large rectangular pulse bias potential,
driving the system well beyond the linear response regime.
As the response of the dot-lead system is highly non-linear, this situation
is quite instructive.  We again study both the $N = 1$ and $N = 2$ cases,
and use the same values for $\epsilon_a$ and $U$ as in the
previous section but with $\Delta=0.2$ meV.
The rectangular pulse bias potential is applied as follows.
At time $t_i = 0.5 \hbar/\Delta $, a sudden upward shift of the bias
with amplitude $\Phi$ is applied to the left lead, leaving the dot energy level
unchanged, and the right lead unbiased as depicted in Fig. \ref{fig1}.  At a
later time $t = t_f$ the bias is turned off.

\subsection{Noninteracting spinless electrons ($N = 1$)}
The abrupt rise in the bias generates a ringing in the response currents
as a result of coherent electronic transitions between
electrons at the Fermi energy in the leads and the dot level.
The period of these oscillations was predicted to be\cite{Wingreenb}:
\begin{equation}
t_p={2 \pi \hbar \over  |\Phi-\epsilon_a|}. \label{period}
\end{equation}
Fig. \ref{fig4} shows, for the case $\epsilon_a = -2 \Delta$,
the response current between the left lead and the dot
upon setting $N=1$ in the $1/N$ expansion.  The oscillation period
accords with Eq. \ref{period}.  For the sake of comparison, Fig. \ref{fig4}
also shows the current for the case of interacting, spinning, electrons.
Note that oscillation period changes when interactions are turned on.
As discussed in the next subsection, this is a consequence of the formation
of Kondo resonances at the Fermi energy in each lead.
It is remarkable that the $1/N$ approach reproduces,
in the $N = 1$ limit, the period of the oscillations expected from Eq. \ref{period}.
Unlike NCA, the $1/N$ approach recovers the
expected features of non-interacting electrons in the $N = 1$ limit.
In particular, many-body Kondo resonances disappear\cite{Gunnarsson}
at $N = 1$ as they must.  As mentioned above, the recovery of
independent-particle physics can be attributed to the
presence of crossing diagrams in the $1/N$ expansion.

Also of interest is the duration of the initial response peak.
The transient response time is shorter in the case of noninteracting
spinless electrons.  The transient response fades away more slowly
because the repulsive Coulomb interaction inhibits
electron motion through the dot.

\subsection{Interacting spinning electrons ($N = 2$)}
In equilibrium it is well known that a Kondo resonance can form
at the Fermi energy of the leads.  For the steady state case of
constant bias potential, previous NCA calculations of spectral density
predicted the splitting of the Kondo peak\cite{Wingreen}
into two resonances, one at the Fermi energy of each lead.
The situation studied here is different as we analyze
the response to a short pulse rather than the steady state behavior.
Nevertheless, the behavior is still consistent with the split Kondo
peak picture.

Fig. \ref{fig5} shows that the response current does not obey Eq. \ref{period}.
Instead the current oscillates with a period that is independent of $\epsilon_a$,
and depends only on the bias.  The period can be seen to be $2 \pi \hbar/\Phi$,
which agrees with Eq. \ref{period} only upon setting $\epsilon_a = 0$.
Fig. \ref{fig6} also demonstrates that the period of the oscillations remains
nearly constant  as the energy of the quantum dot level is
varied from $\epsilon_a=0$ (mixed-valence regime) to $\epsilon_a = -3 \Delta$
(Kondo regime).  Large biases split the Kondo peak
in two, and the resulting many-body resonances are separated by energy
$\Phi$. Electronic transitions between these two peaks induce
oscillations of period $2 \pi \hbar/\Phi$.
The $1/N$ calculation agrees qualitatively with the NCA calculation of
Plihal, Langreth, and Nordlander\cite{Plihal}.  Figs. \ref{fig5} and \ref{fig6}
show that the split-peak interpretation holds even for very large biases,
$\Phi = 2 \Delta$ to $10 \Delta$, and also in the mixed-valence regime.
Fig. \ref{fig6} also shows that the magnitude of the response current
gradually decreases, and the transient oscillations damp out more quickly,
as the dot energy level is moved downwards in energy.  Insight into
this behavior may again be attained from study of a quantum dot in equilibrium.
In the limit $\epsilon_a/\Delta \rightarrow -\infty$
the Kondo scale vanishes and the spectral weight at the Fermi
energy is suppressed.  The dot enters the Coulomb blockade regime,
inhibiting the flow of current.

Finally we discuss the effect of non-zero temperature on the response
currents. Fig. \ref{fig7} shows, for both noninteracting spinless and interacting
spinning electrons, the currents at three temperatures: $T=100$ mK, $T=300$ mK
and $500$ mK. The latter temperature is sufficiently high that multiple particle-hole
pairs will be excited, as the level spacing is only $D/(13 M) \approx 120$ mK at the Fermi
energy; therefore the Hilbert space restriction to at most two particle-hole pairs is a severe limitation.  
Nevertheless, there is little change in the response of the spinless electrons
as the temperature is increased.  This is as it should be since the temperatures are still well below the
Fermi temperature of the leads.  By contrast the $N = 2$ case shows a significant
decrease in the current magnitude, and especially the oscillations, at
the higher temperatures.  Using the finite-$U$ formula Eq. \ref{Hal}
with $U / \Delta = 10$, yields $T_K \approx 600$ mK.  Considering the uncertainities
involved in estimating the Kondo temperature (Eq. \ref{Hal} gives the order of magnitude
but the exact multiplicative factor is unknown) we can expect a suppression of 
the Kondo resonance and its associated effects at $500$ mK.  The behavior of the currents shown in
Fig. \ref{fig7} is consistent
with this interpretation, as the period of the transient oscillations decreases
to approach that of the spinless system.

\section{Conclusions}
\label{sec:concl}

We have presented a method for calculating the
response of a quantum dot to time-dependent bias potentials. The
approach, which is based upon a truncation of the Hilbert space, is
systematic because corrections can be incorporated by including
higher powers in the $1/N$ expansion.
In agreement with previous approaches we find coherent
oscillations in the response currents at low temperatures. We note
that although the frequency of these oscillations is of order
a terahertz for parameters typical of quantum dot devices, it should be
possible to detect the oscillations experimentally\cite{Nakamura}.
The dynamical $1/N$ approach permits a realistic
description of the quantum dot as finite Coulomb repulsion and multiple
dot levels may be modeled.  For typical device parameters,
response currents are qualitatively
similar to those found in the $U \rightarrow \infty$ limit. However,
the magnitude of the current is larger.

We also discussed an extension of the dynamical $1/N$ expansion
to treat the case of non-zero temperature.  In the case of interacting spinning
electrons the magnitude of the currents, and the nature of the
transient response are sensitive to relative size of the system temperature in
comparison to the Kondo temperature.  In contrast non-interacting spinless
electrons show little temperature dependence.

The dynamical $1/N$ approach complements other methods such as NCA and
the TDMRG algorithm.  The primary limitation
of the dynamical $1/N$ approach comes from the truncation of the Hilbert space
in which only a small number of particle-hole pairs (in this paper, at most two)
are permitted. Hence, as it stands the method is not well suited for the study of
steady state situations, and we cannot use it, for instance, to answer the
question of whether or not a lead-dot-lead system exhibits coherent oscillations at
long times\cite{Coleman,Rosch}.  As NCA sums up an infinite set of Feynman
diagrams, including ones with arbitrary numbers of particle-hole pairs, it provides
a far better description of long-time behavior, including steady state current flow.
For the same reason NCA is also superior at high temperature.
However finite Coulomb interactions and other realistic features of lead-dot-lead
systems are technically difficult to incorporate within NCA, but not in the dynamical
$1/N$ method.  Also, NCA shows unphysical behavior in the mixed valence regime,
at very low temperatures, and in the spinless $N = 1$ limit.
The dynamical $1/N$ approach does not suffer from
these pathologies.   The method shares some features with the TDMRG
algorithm as both are real-time approaches that truncate the Hilbert space
in a systematic, though in a different, fashion.  Unlike the present approach, TDMRG
can treat strong electron interactions between electrons in the leads and can be used to
study tunneling between Luttinger liquids\cite{Cazalilla}.  
On the other hand, non-interacting electrons in the leads are treated exactly 
within the dynamical $1/N$ approach.  In TDMRG these require as much or 
more computational effort as interacting electrons.

A complete description of the electronic transport properties in a
SET would require taking into account the whole energy spectrum of
the quantum dot and would include both the direct Coulomb repulsion and
spin-exchange between electrons in different energy levels of the
dot.  Also the conducting leads should be described by realistic
densities of states.  The dynamical $1/N$ approach is well suited to
accommodate these complications.  Another aspect worth further attention
is the case of orbital degeneracy of the dot levels.  As the degeneracy increases,
the Kondo resonance strengthens, and its width decreases\cite{Kroha}.
A combined experimental and theoretical exploration of the
effect of orbital degeneracy on transport properties would be interesting.
Finally, it should also be possible to extend the method to describe
two coupled quantum dots attached to leads\cite{Aguado}.
Competition between Kondo resonances and spin-exchange between the two
dots leads to unusual features such as a non-Fermi liquid
fixed point\cite{Jones,Affleck}.  Probing nonlinear transport 
as the parameters are tuned through this fixed point\cite{Georges}
could possibly shed some light on the physics of heavy-fermions materials.

\acknowledgments

We thank R. Aguado, M. A. Cazalilla,  J. C. Cuevas, O. Gunnarsson,
R. L\'opez, R. H. McKenzie, and J. Weis for very helpful
discussions and K. Held for a careful reading of the manuscript.
J. M. was supported by the European Community under fellowship
HPMF-CT-2000-00870. J. B. M. was supported by the NSF under grant
Nos. DMR-9712391 and DMR-0213818.
Calculations were carried out in double-precision C
on an IBM SP4 machine at the
Max-Planck-Institut. Parts of this work were carried out at
the University of Queensland.

\section{Appendix}
\label{sec:appendix}

In this Appendix we present details of the equations of motion and
the calculation of the currents.

\subsection{Equations of Motion}
Equations of motion appropriate for the atom-surface scattering problem have
been published elsewhere\cite{Onufriev}.  We rewrite them here,
generalizing the equations to include two sectors that are
of order O($1/N^2$) and substituting two conducting leads for the metallic surface.
To remove diagonal terms in the equations of motion we introduce the phase factor
\begin{equation}
\phi^{(i)}_a(t) \equiv {{1}\over{\hbar}}~ \int^t_0 \epsilon^{(i)}_a(t^\prime) dt^\prime\ ,
\end{equation}
which is the phase of the quantum dot level when it is decoupled from the leads, and
\begin{equation}
\phi_{k \alpha}(t) \equiv {{1}\over{\hbar}}~ \int^t_0 \epsilon_{k \alpha}(t^\prime) dt^\prime\ ,
\end{equation}
the phase of the electronic levels in the decoupled leads.  Upon projecting the
time-dependent
Schr\"odinger equation onto the different sectors of the Hilbert space,
the equations of motion may then be extracted.  Following Refs. \onlinecite{Ernie}
and \onlinecite{Onufriev} we use capital letters to denote amplitudes in which
the diagonal phases of the corresponding lower case amplitudes have been factored out:

\begin{eqnarray}
i \hbar {d\over{dt}} F &=& \sum_{a; k \alpha} V^{(1)*}_{a; k \alpha}
\exp \{i[\phi_{k \alpha}(t) - \phi^{(1)}_a(t)]\}\ B_{a; k \alpha}
\nonumber\\
i \hbar {d\over{dt}} B_{a; k \alpha} &=& V^{(1)}_{a; k \alpha}\
\exp \{i[\phi^{(1)}_a(t) - \phi_{k \alpha}(t)]\}\ F
\nonumber \\
&+& \delta_{a,0}~ \sqrt{1 - 1/N} \sum_{q, \beta} V^{(2)*}_{0; q \beta}
\exp \{-i[U - \phi_{q \beta}(t) + 2 \phi^{(2)}_0(t) - \phi^{(1)}_0(t)] \}\
[\theta(k-q)~ D_{k \alpha, q \beta} + \theta(q-k)~ D_{q \beta, k \alpha}]
\nonumber\\
&+& {{1}\over{\sqrt{N}}} \sum_{L \gamma} V^{(1)}_{a; L \gamma}\
\exp \{i[\phi^{(1)}_a(t) - \phi_{L \gamma}(t)]\}\ E_{L \gamma, k \alpha}
\nonumber\\
i \hbar {d\over{dt}}E_{L \gamma k \alpha} &=& {{1}\over{\sqrt{N}}} \sum_a V^{(1)*}_{a; L \gamma}\
\exp \{i[\phi_{L\gamma}(t) - \phi^{(1)}_a(t)]\}\ B_{a; k \alpha}
\nonumber\\
&+& \sqrt{{{N - 1}\over{2N}}} \sum_{a; q \beta} V^{(1)*}_{a; q \beta}
\exp \{i[\phi_{q\beta}(t)- \phi^{(1)}_a(t)]\}
[\theta(k-q)\ S_{a; L \gamma, k \alpha, q \beta} + \theta(q-k)\ S_{a; L \gamma, q \beta, k \alpha}]
\nonumber\\
&+& \sqrt{{{N + 1}\over{2N}}}
\sum_{a;q\beta}V^{(1)*}_{a; q \beta} \exp \{i[\phi_{q\beta}(t) - \phi^{(1)}_a(t)]\}
[\theta(k-q)\ A_{a; L \gamma, k \alpha, q \beta} -
\theta(q-k)\ A_{a;L \gamma, q \beta, k \alpha}]
\nonumber\\
i \hbar {d\over{dt}} D_{k \alpha, q \beta} &=& \sqrt{1 - 1/N} \
V^{(2)}_{0; q \beta}\
\exp \{i[U - \phi_{q \beta}(t) + 2 \phi^{(2)}_0(t)- \phi^{(1)}_0(t)]\}\ B_{0; k \alpha}
\nonumber\\
&+& \sqrt{1 - 1/N}~ V^{(2)}_{0; k \alpha}~
\exp \{i[U - \phi_{k \alpha}(t) + 2 \phi^{(2)}_0(t)- \phi^{(1)}_0(t)]\}~ B_{0; q \beta}
\nonumber\\
&+& \sqrt{{{2}\over{N}}} \sum_{L \gamma} V^{(2)}_{0; L \gamma}\
\exp \{i[U - \phi_{L \gamma}(t) + 2 \phi^{(2)}_0(t) - \phi^{(1)}_0(t)]\}~
S_{0;L \gamma, k \alpha, q \beta}
\nonumber\\
i \hbar {d\over{dt}} S_{a; L \gamma, k \alpha, q \beta} &=&
\delta_{a,0}\ \sqrt{{{2}\over{N}}}\ V^{(2)*}_{0; L \gamma}\
\exp \{-i[U - \phi_{L \gamma}+ 2 \phi^{(2)}_0(t) - \phi^{(1)}_0(t)]\} D_{k \alpha, q \beta}
\nonumber\\
&+& \sqrt{{{N - 1}\over{2N}}}~ [V^{(1)}_{a; q \beta}
\exp \{i[\phi^{(1)}_a(t) - \phi_{q \beta}(t)] \}\ E_{L \gamma, k \alpha} +
V^{(1)}_{a; k \alpha} \exp \{i[\phi^{(1)}_a(t) - \phi_{k \alpha}(t)]\} \
E_{L \gamma, q \beta}]\
\nonumber\\
&+& {{1}\over{\sqrt{N}}} \sum_{J \delta}
V^{(1)*}_{a; J \delta} \exp \{i[\phi_{J\delta}(t) - \phi^{(1)}_a(t)]\}
[\theta(L-J) G_{a; L \gamma, J \delta, k \alpha, q \beta}
+ \theta(J-L) G_{a; J \delta, L \gamma, k \alpha, q \beta}]
\nonumber\\
i \hbar {d\over{dt}} A_{a; L \gamma, k \alpha, q \beta} &=& \sqrt{{{N + 1}\over{2N}}}~
[V^{(1)}_{a; q \beta} \exp\{i[\phi^{(1)}_a(t) - \phi_{q \beta}(t)]\}\
E_{L \gamma, k \alpha} - V^{(1)}_{a; k \alpha} \exp \{i[\phi^{(1)}_a(t) -
\phi_{k \alpha}(t)]\}\ E_{L \gamma, q \beta}]\
\nonumber\\
&+& {{1}\over{\sqrt{N}}} \sum_{J \delta}
V^{(1)*}_{a; J \delta} \exp \{i[\phi_{J \delta}(t)- \phi^{(1)}_a(t)]\}
[\theta(L-J) H_{a; L \gamma, J \delta, k \alpha, q \beta}
-\theta(J-L) H_{a; J \delta, L \gamma, k \alpha, q \beta}]
\nonumber\\
i \hbar {d\over{dt}} G_{L \gamma, J \delta, k \alpha, q \beta}
&=& {{1}\over{\sqrt{N}}} V^{(1)*}_{a; J \delta}
\exp \{i[\phi_{J \delta}(t)-\phi^{(1)}_a(t)]\} S_{a; L \gamma, k \alpha, q \beta} +
{{1}\over{\sqrt{N}}} V^{(1)*}_{a; L \gamma} \exp \{i[\phi_{L \gamma}(t)-\phi^{(1)}_a(t)]\}
S_{a; J \delta, k \alpha, q \beta}
\nonumber\\
i \hbar {d\over{dt}} H_{L \gamma, J \delta, k \alpha, q \beta}
&=& {{1}\over{\sqrt{N}}} V^{(1)*}_{a; J \delta}
\exp \{i[\phi_{J \delta}(t)-\phi^{(1)}_a(t)]\} A_{a; L \gamma, k \alpha, q \beta} -
{{1}\over{\sqrt{N}}} V^{(1)*}_{a; L \gamma} \exp \{i[\phi_{L \gamma}(t)-\phi^{(1)}_a(t)]\}
A_{a; J \delta, k \alpha, q \beta}\ .
\label{EOM}
\end{eqnarray}
Here $\theta(k - q)$ and similar terms are unit step functions that enforces the ordering convention
on the continuum indices.  The equations of motion are numerically integrated forward in time
using a fourth-order Runge-Kutta algorithm with adaptive
time steps.  We monitor the normalization of the wavefunction to ensure that any departure
from unitary evolution remains small, less than $10^{-6}$.

\subsection{Calculation of Currents}
\begin{figure}
\includegraphics[width=18cm,height=14cm]{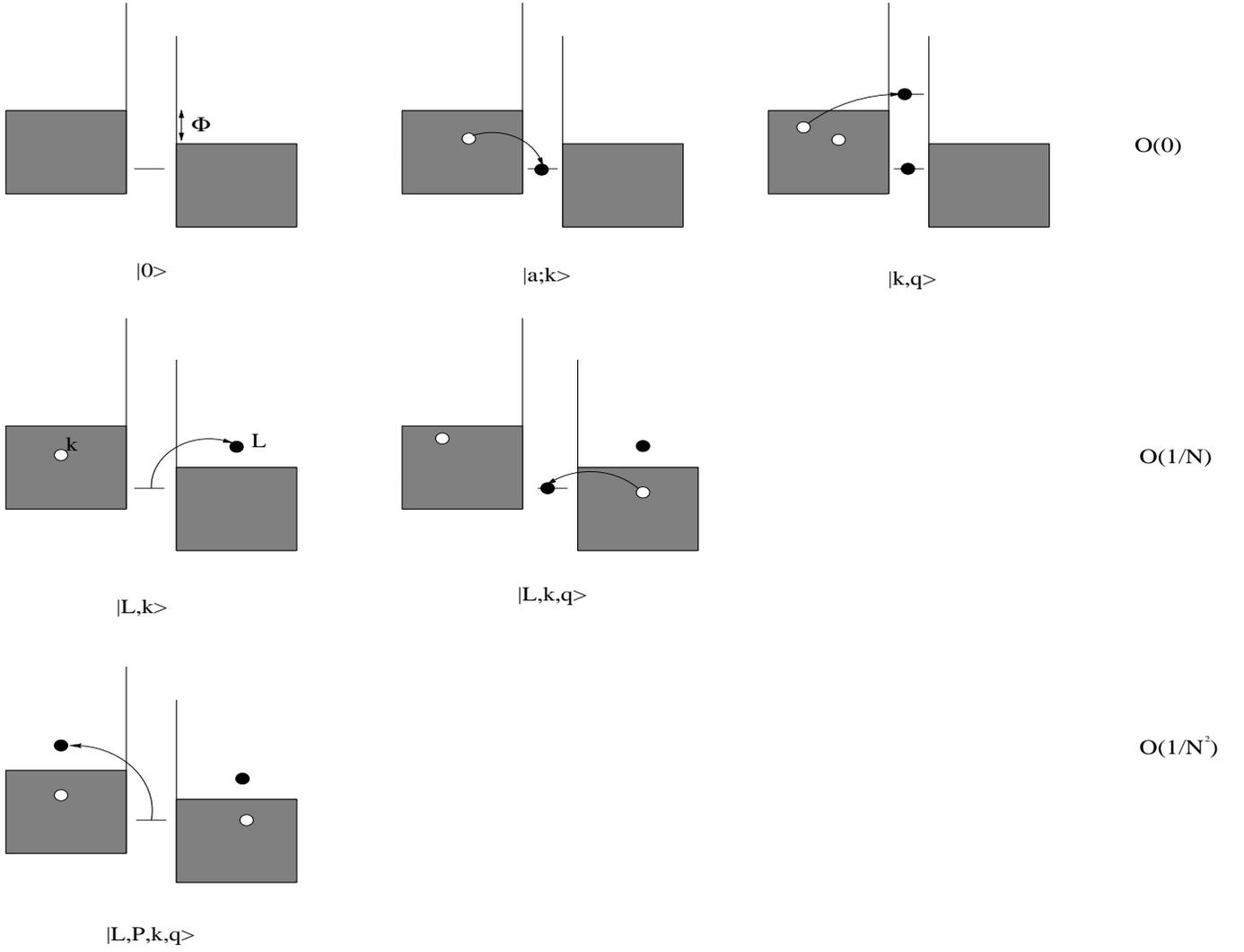}
\caption{\label{fig2}
Schematic representation of Hilbert space as organized by
the $1/N$ expansion. The diagram shows the succesive
electronic transitions due both to the coupling of the dot
levels to the leads and the applied bias.  Particle-hole excitations are
produced in both leads. Sectors with increasing number of particle-hole pairs
appear at successive orders in the $1/N$ expansion.
We retain configurations which are at most of order O($1/N^2$).
Not shown, but included in the many-body wavefunction, are configurations
with two particle-hole pairs in a single lead, or two particles in one lead and
two holes in the other lead.
}

\end{figure}

\begin{figure}
\center
\includegraphics[width=10cm,height=10cm]{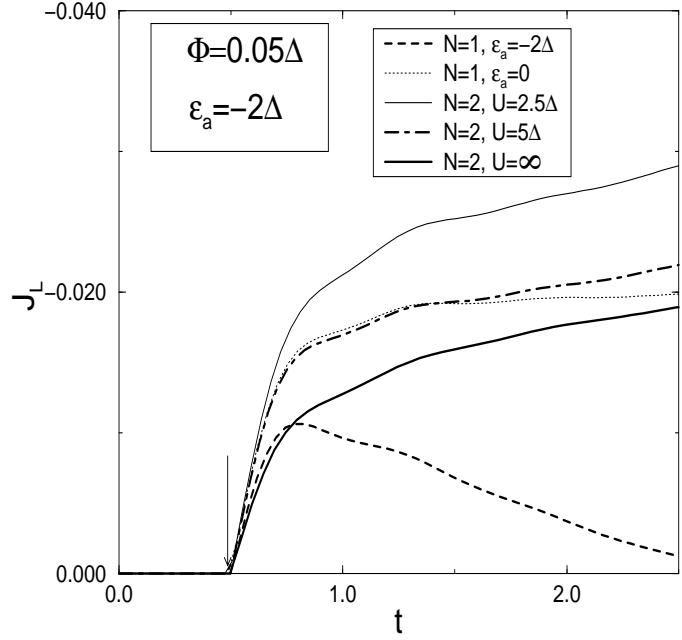}
\caption{\label{fig3} Response of the electric current per spin
channel between the left lead and a quantum dot to a small
symmetric bias step potential.  Time is given in units of
$\hbar/\Delta$ where $\Delta=0.4$ meV, and current is in units of
$e/h$.  A small, symmetric, step bias potential of amplitude
$\Phi/2 = 0.025 \Delta$ is applied to the left lead and an
opposing bias of $-0.025 \Delta$ is applied to the right lead. 
The arrow at time $t = t_i = 0.5 \hbar/\Delta $ indicates the moment 
when the step in the bias is turned on.  
The electronic level in the dot is held fixed at energy
$\epsilon_a = -2 \Delta$.  (For comparison in the $N=1$ case we also show results  
for $\epsilon_a = 0$.) 
The non-interacting spinless case as calculated within
the dynamical $1/N$ approximation upon setting $N=1$ is compared
to the case of interacting electrons ($N = 2$) with $U = 1$ and $2$ meV
and $U = \infty$. Currents initially grow and then
decay (evident in the time range shown here only for the case $N=1$ and $\epsilon_a=-2 \Delta$) 
despite the fact that the bias remains on. 
The eventual decay of the currents is a consequence of continued production of 
particle-hole pairs, saturating the high-order sectors of the $1/N$ expansion.  
The response current is strongly enhanced in the interacting case
with respect to the spinless case due to the presence of a Kondo
resonance at the Fermi level of the leads. For $U = 1$ and $2$ meV, the
current is further enhanced reflecting an increase in the Kondo
temperature.}
\end{figure}

\begin{figure}
\center
\includegraphics[width=10cm,height=10cm]{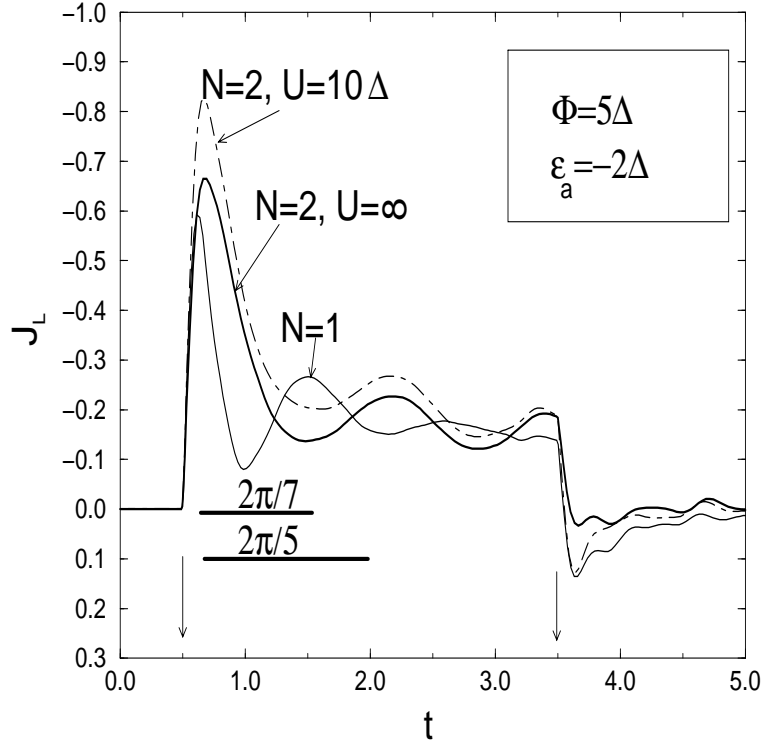}
\caption{\label{fig4} Electric current per spin channel between
the left lead and a quantum dot due to application of a large rectangular pulse
bias potential.
Time is given in units of $\hbar/\Delta$ where $\Delta=0.2$ meV,
and current is in units of $e/h$. 
Here the bias of amplitude $\Phi = 5 \Delta$ 
is applied only to the left lead (see Fig. \ref{fig1}), and the
level in the dot is held fixed at $\epsilon_a = -2 \Delta$. The bias
is turned on abruptly at time
$t = 0.5 \hbar/\Delta$ and off at time $t=3.5 \hbar/\Delta$.
In the same plot we compare the non-interacting,
spinless case ($N=1$) with the $N=2$ case for both the $U \rightarrow \infty$
limit and for the experimentally relevant value of $U=2$
meV of a SET\cite{Goldhaber}.  The transient response time of the interacting electrons
is longer than for non-interacting electrons.
As can be seen upon comparison with the time scales (horizontal lines)
the period of the oscillations also increases as $N$ changes from 1 to 2,
in accord with the formula $t_p= {2 \pi \hbar \over |\Phi - \epsilon_a|}$ but only
if we set $\epsilon_a = 0$ in the case of spinning electrons.}
\end{figure}

\begin{figure}
\center
\includegraphics[width=15cm,height=15cm]{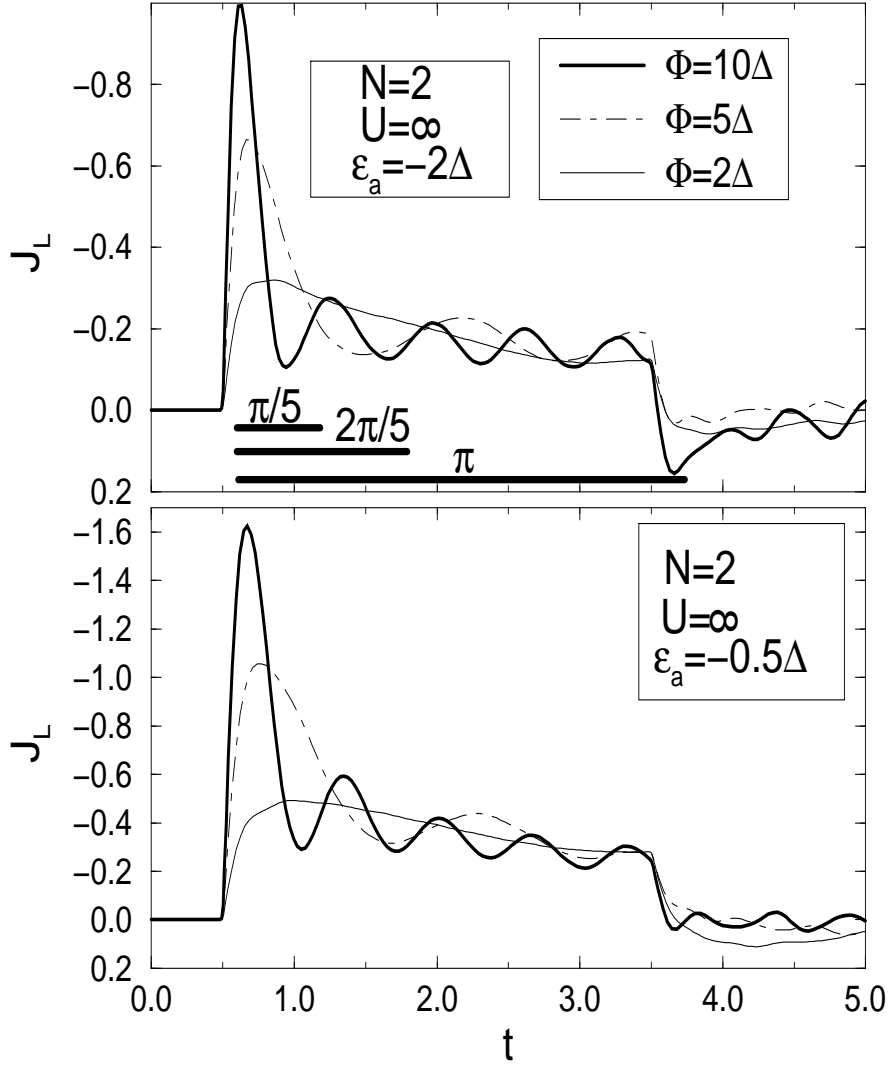}
\caption{\label{fig5} Dependence of the electronic current per
spin channel between the left lead and the quantum dot on the
amplitude, $\Phi$, of the externally applied rectangular pulse
bias potential (see Fig. \ref{fig1}).
Time is in units of $\hbar/\Delta$ where $\Delta = 0.2$ meV, and
current is in units of $e/h$.
We consider interacting spinning electrons ($N = 2$)
and take $U \rightarrow \infty$.  The quantum dot level is either in the
Kondo regime ($\epsilon_a/\Delta=-2$) or in the mixed valent
regime ($\epsilon_a/\Delta=-0.5$). The pulse starts at time
$t=0.5 \hbar/\Delta$ and ends at time $t=3.5 \hbar/\Delta$.
The current is plotted for different applied biases:
$\Phi = 2 \Delta, 5 \Delta,$ and $10 \Delta$. A comparison
of time scales (horizontal lines) shows that the period
of the oscillations does not obey Eq. \ref{period} but rather
$t_p = 2 \pi \hbar / \Phi$.}
\end{figure}

\begin{figure}
\center
\includegraphics[width=10cm,height=10cm]{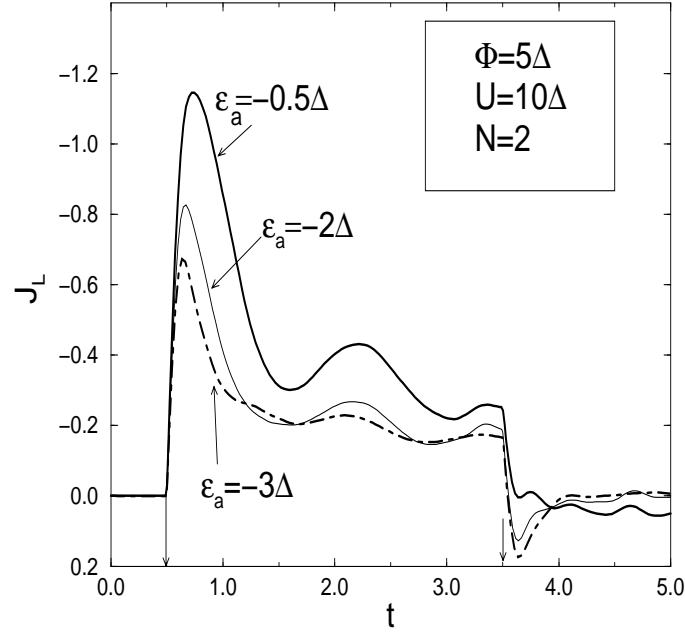}
\caption{\label{fig6} Electric current per spin channel between
the left lead and the quantum dot for different dot level energies.
Time is in units of $\hbar/\Delta$ where $\Delta=0.2$ meV,
and current is in units of $e/h$.  We fix $N =2$ and $U=2$ meV,
values appropriate for a semiconducting quantum dot.
A rectangular pulse bias potential of amplitude $\Phi = 5 \Delta$ is
applied with the dot level set at: $\epsilon_a = -3 \Delta$, $-2 \Delta$,
and $-0.5 \Delta$.
The pulse is abruptly turned on at time $t=0.5 \hbar/\Delta$ and off at time
$t=3.5 \hbar/\Delta$ as marked by the arrows. We find that the period of
the oscillations remains constant even as the quantum dot level is moved
from the Kondo into the mixed-valence regime. As $\epsilon_a/\Delta$
becomes more negative the Kondo scale decreases and the magnitude of the current
is also suppressed. }
\end{figure}

\begin{figure}
\center
\includegraphics[width=15cm,height=15cm]{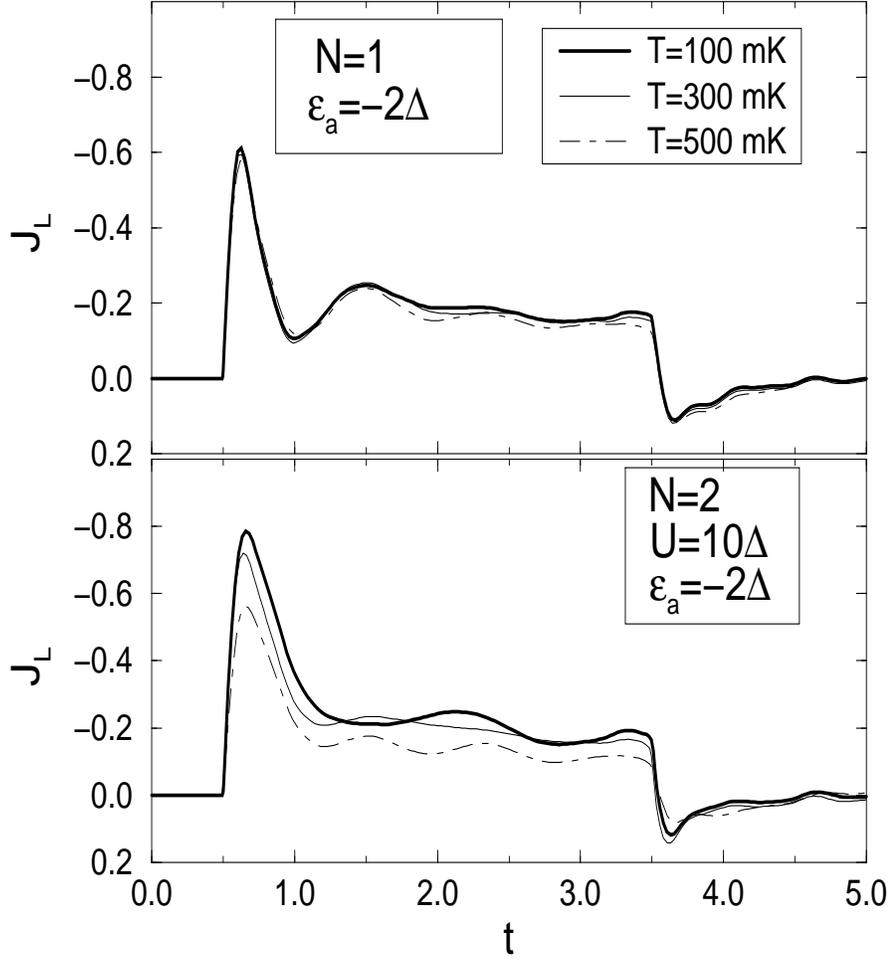}
\caption{\label{fig7} Temperature dependence of the response
current per spin channel through a quantum dot. Noninteracting spinless
($N=1$) electrons are compared to interacting spinful electrons ($N=2$).
Time is in units of $\hbar/\Delta$ where $\Delta = 0.2$ meV,
and current is in units of $e/h$.  
A rectangular pulse bias potential of amplitude $\Phi = 5 \Delta$ is applied.
We set $\epsilon_a = -2\Delta$, and for the interacting case we set $U=2$ meV. 
In contrast to the noninteracting spinless case,
for interacting spinful electrons the period of the oscillations and the
magnitude of the current decrease significantly
as the temperature is increased, approaching those of the
spinless system.}
\end{figure}

A consideration of the rate of change of the quantum dot occupancy, as
determined from the commutator $[n_a,~ H]$, permits the straightforward identification
of the current operators between the dot and the left and right leads:
\begin{eqnarray}
J_{\alpha} &=&  {{e}\over{\hbar \sqrt{N}}} \sum_{a;k} \bigg{\{} i [V^{(1)*}_{a; k\alpha} \hat P_1
+ V^{(2)*}_{a; k \alpha} \hat P_2 ] c_{k\alpha}^{\dag \sigma} c_{a \sigma}
 + H.c  \bigg{\}}
\nonumber\\
&=& -{{2e}\over{\hbar \sqrt{N}}} {\rm Im} \bigg{\{} \sum_{a;k} [V^{(1)*}_{a; k\alpha} \hat P_1 +
 V^{(2)*}_{a; k\alpha} \hat P_2] c_{k\alpha}^{\dag \sigma} c_{a \sigma}
\bigg{\}}
\label{currentexp}
\end{eqnarray}
where as before the Greek index $\alpha$ refers to either the left or right lead.
The expectation value of the current is calculated as the quantum and thermal average of the
operator Eq. \ref{currentexp}.  For each time-evolved
wavefunctions, $\tilde{\Psi}_n(t)$, the corresponding expected current may be
expressed in terms of the amplitudes in the different sectors:
\begin{eqnarray}
\lefteqn{ \langle \tilde{\Psi}_n(t) | J_{\alpha} | \tilde{\Psi}_n(t) \rangle
= -{{2 e}\over{\hbar}}~ {\rm Im} \bigg{\{} \sum_{a; k} V^{(1)*}_{a; k \alpha}
\exp \{ i[\phi_{k \alpha}(t) - \phi^{(1)}_a(t)] \} F^* B_{a; k \alpha}}
&&
\nonumber\\
&+& \sqrt{1 - 1/N} \sum_{k, q \beta}  V^{(2)*}_{0; k \alpha}
\exp \{-i[U t -\phi_{k \alpha}(t) + 2 \phi^{(2)}_0(t) - \phi^{(1)}_0(t)] \}
B^*_{0; q \beta} [\theta(k-q) D_{k \alpha, q \beta} + \theta(q-k) D_{q \beta, k \alpha}] 
\nonumber\\
&+& {{1}\over{\sqrt{N}}} \sum_{a; L, q \beta} V^{(1)*}_{a; L \alpha}
\exp \{ i[\phi_{L \alpha}(t) - \phi^{(1)}_a(t)] \} E^*_{L \alpha; q \beta} B_{a; q \beta}
\nonumber\\
&+& \sqrt{{{N-1}\over{2N}}} \sum_{a; q, L \gamma, k \beta} V^{(1)*}_{a; q \alpha}
\exp \{ i[\phi_{q \alpha}(t) - \phi^{(1)}_a(t)] \}
E^*_{L \gamma, k \beta} [\theta(k-q) S_{a; L \gamma, k \beta, q \alpha}
+ \theta(q-k) S_{a; L \gamma, q \alpha, k \beta}]
\nonumber\\
&+& \sqrt{{{N+1}\over{2N}}} \sum_{a; q, L \gamma, k \beta} V^{(1)*}_{a; q \alpha}
\exp \{ i[\phi_{q \alpha}(t) - \phi^{(1)}_a(t)] \}
E^*_{L \gamma, q \alpha}[\theta(k-q) A_{a; L \gamma, k \beta, q \alpha}-
\theta(q-k) A_{a; L \gamma, q \alpha, k \beta}]
\nonumber\\
&+& \sqrt{{{2}\over{N}}} \sum_{L, k \gamma, q \beta} V^{(2)*}_{0; L \alpha}
\exp \{ -i[U t - \phi_{k \gamma}(t)+ 2\phi^{(2)}_0(t) -
\phi^{(1)}_0(t)] \} S^*_{0; L \alpha k \gamma, q \beta} D_{k \gamma; q \beta}
\nonumber\\
&+& {{1}\over{\sqrt{N}}} \sum_{a; L, J \beta, k \gamma, q \delta} V^{(1)*}_{a; L \alpha}
\exp \{ i[\phi_{J \beta}(t) - \phi^{(1)}_a(t)] \}
S_{a; J \beta, k \gamma, q \delta} [\theta(L-J) G^*_{L \alpha, J \beta, k \gamma, q \delta}
+ \theta(J-L) G^*_{J \beta, L \alpha, k \gamma, q \delta}]
\nonumber\\
&+& {{1}\over{\sqrt{N}}} \sum_{a; L, J \beta, k \gamma, q \delta} V^{(1)*}_{a; L \alpha}
\exp \{ i[\phi_{J \beta}(t) - \phi^{(1)}_a(t)] \}
A_{a; J \beta, k \gamma, q \delta} [\theta(L-J) H^*_{L \alpha, J \beta, k \gamma, q \delta}
- \theta(J-L) H^*_{J \beta, L \alpha, k \gamma, q \delta}]
\bigg{\}}
\label{current}
\end{eqnarray}
Note that the current remains finite in the $N \rightarrow \infty$ limit as the
amplitudes appearing in Eq. \ref{current} are at most O(1).  Thus Eq. \ref{current}
is the total current summed over all $N$ spin channels and divided by a factor of $N$,
and hence equivalent to the current per spin channel (since spin-rotational invariance
remains unbroken).  Fig. \ref{fig2} depicts some of the contributions to the current between
the dot and the leads.  At each time step we check that current is conserved
to within an accuracy of $10^{-6}$ (when charging of the dot is taken into account).
A thermal average of the current operator, using Eq. \ref{aver}, is the final step:
\begin{equation}
\langle J_\alpha(t) \rangle= {1 \over Z} \sum_n \langle
\tilde{\Psi}_n(t) | J_{\alpha} | \tilde{\Psi}_n(t) \rangle.
\end{equation}

\newpage

\end{document}